\documentclass[a4paper,12pt]{article}
\usepackage{amsmath}
\usepackage{graphicx}
\author{Oscar Stiffelman\\
    \texttt{ozzie@cs.stanford.edu}\\
    \texttt{imagine@gmail.com}
}

\title{PivotCompress: Compression by Sorting}

\begin{document}

\maketitle

\begin{abstract}
Sorted data is usually easier to compress than unsorted permutations
of the same data.  This motivates a simple compression scheme: specify
the sorted permutation of the data along with a representation of the
sorted data compressed recursively.  The sorted permutation can be
specified by recording the decisions made by quicksort.  If the size
of the data is known, then the quicksort decisions describe the data
at a rate that is nearly as efficient as the minimal prefix-free code
for the distribution, which is bounded by the entropy of the
distribution.  This is possible even though the distribution is
unknown ahead of time.  Used in this way, quicksort acts as a
universal code in that it is asymptotically optimal for any stationary
source.  The Shannon entropy is a lower bound when describing
stochastic, independent symbols.  However, it is possible to encode
non-uniform, finite strings below the entropy of the sample
distribution by also encoding symbol counts because the values in the
sequence are no longer independent once the counts are known.  The key
insight is that sparse quicksort comparison vectors can also be
compressed to achieve an even lower rate when data is highly
non-uniform while incurring only a modest penalty when data is random.

\end{abstract}

Sorted data is usually easier to compress than unsorted permutations
of the same data.  Because sequential values are in order when data is
sorted, their non-negative differences can be encoded in place of the
original values.  And because repeated values are all contiguous, they
can be encoded by including a count along with the first
instance. This motivates a simple compression scheme: first describe
how to permute the data into sorted order, and then describe the
sorted data.  The permutation is invertible, so it can be used
along with the description of the sorted data to generate the
original data.

\begin{figure}[p]
    \includegraphics[scale=0.5]{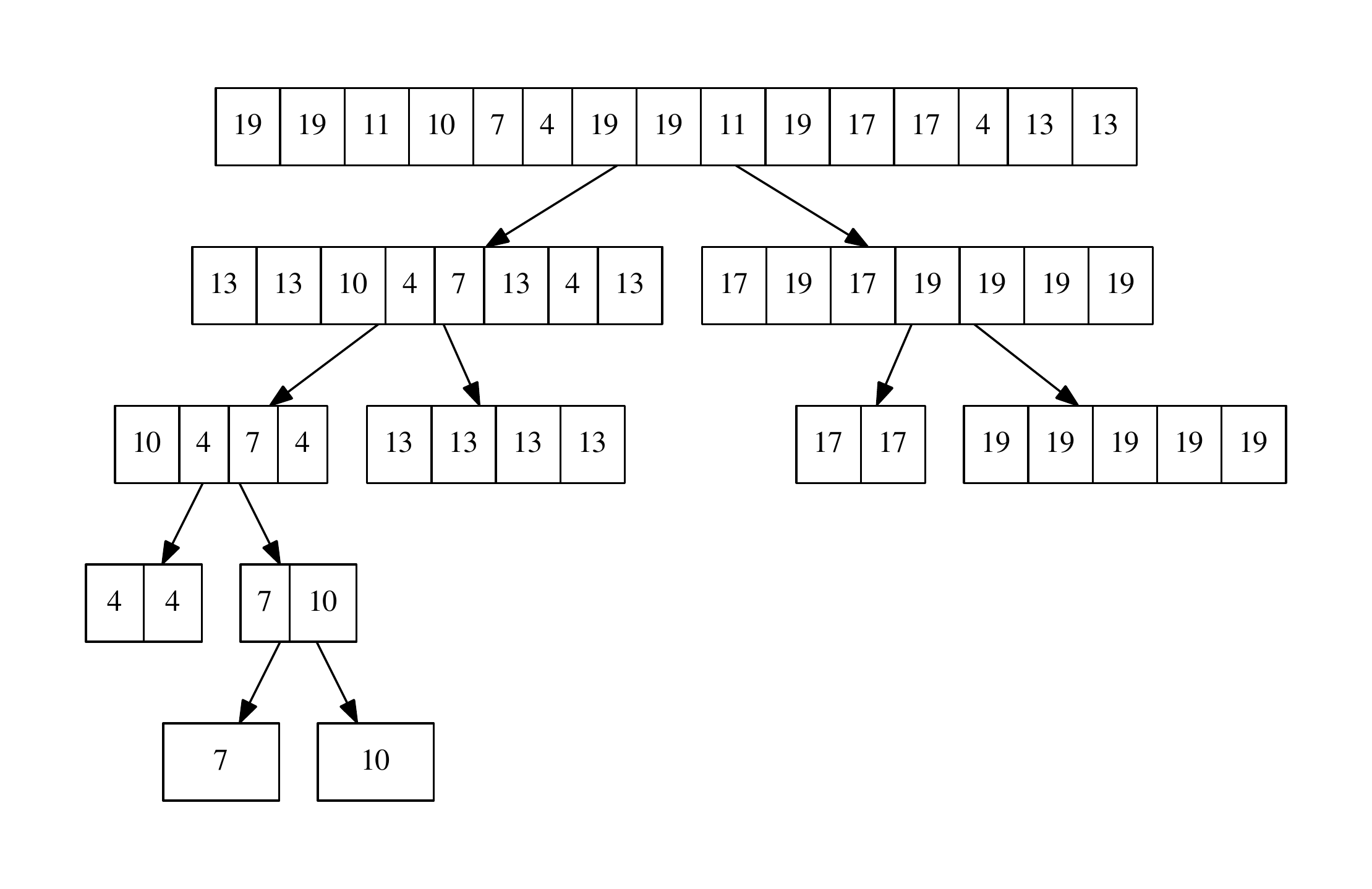}
    \caption{quicksort partition tree}
    \label{fig:quicksort}
\end{figure}

\begin{figure}[p]
    \includegraphics[scale=0.5]{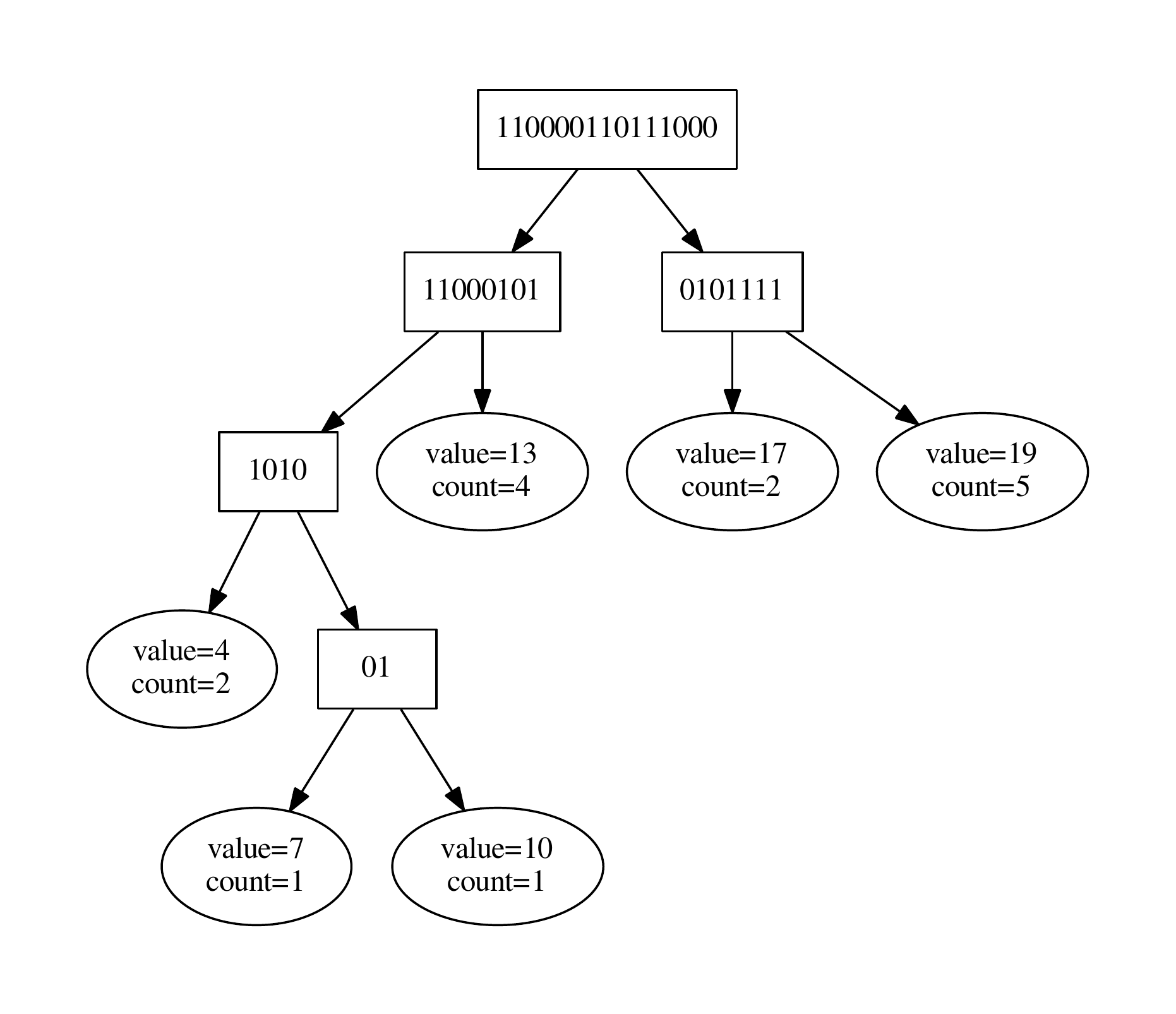}
    \caption{decision bitvectors}
    \label{fig:decisionvectors}
\end{figure}

Because there are $N!$ permutations of a sequence of $N$ items,
$\log_2(N!)$ bits are required to specify a particular permutation.
But if there is any frequency regularity in the data, it is possible
to specify the sorted permutation using fewer bits because there are
fewer distinct permutations.  One way to specify the sorted
permutation of data that uses fewer bits when data is non-uniform is
to record the decisions made by the quicksort algorithm as it
recursively partitions the data around pivots.  In quicksort, each
item in the sequence is compared to a pivot value.  Based on that
comparison, it is assigned to either the left or the right partition,
and the algorithm is applied recursively to each partition.  The
recursion terminates when the sequence cannot be partitioned, either
because it is a single item or because all of the items are equal.
When quicksort terminates, all of the occurrences of a given symbol
are associated with a single leaf node, and the leaves are arranged in
sort order~\cite{cormen}.  Even without the original data and pivot
values, the sorted sequence can be regenerated by running the
quicksort algorithm again, using the recorded decisions instead of
comparing data to pivots.  Using the recorded decisions, the index
array $X[i] = i$ is transformed by this algorithm into a permutation
vector $X[i]=j$, indicating that the value in position $i$ is moved to
position $j$.  The inverse permutation vector, defined by $Y[X[i]] =
i$ can be used to permute the sorted data back into the original
order.  An example quicksort partition tree is shown in
figure~\ref{fig:quicksort}.  The corresponding decision bitvectors are
depicted in figure~\ref{fig:decisionvectors}.

Assume the initial sequence is of length $N$, and there are $M$ unique
symbols. Let $c_i$ represent the count of the $i$'th symbol.  When the
pivots are selected so that the partitions are maximally uniform, the
leaf for the $i$'th symbol is reached after approximately
$\log_2(N/c_i)$ partitions.  Even if the data cannot be uniformly
partitioned, at most one extra partition is necessary.  For each of
the $c_i$ instances of a given symbol, all of the comparisons and
corresponding partition decisions need to be recorded.  The total
number of comparisons is approximately $$\sum{c_i*\log_2(N/c_i)}.$$
Using $p_i=c_i/N$, this can be written as the familiar
$$N*\sum{p_i*\log_2(1/p_i)},$$ showing that the number of comparisons
on average is approximately equal to the entropy of the distribution.
The maximally uniform quicksort tree is in fact equivalent to a Fano
prefix code tree for the distribution, which is no more than one bit
worse than the minimal Huffman code tree~\cite[P.123]{Cover}.  If the
comparisons are recorded using bitvectors at each node, with each
comparison stored in a single bit, then the description of the sorted
permutation is also a description of the data that is optimal or near
optimal for the distribution.  Interestingly, this description
efficiency is possible even though the distribution is unknown ahead
of time and does not need to be encoded before the data.  The only
information required to decode it is the number of elements $N$, which
can be encoded using a universal code in $2*\log_2(N)+1$
bits~\cite{elias}.  This means that quicksort acts as a universal code
in that it is asymptotically optimal for any stationary source.

Perhaps surprisingly, when the distribution is nonuniform, it is
possible to encode the data using even fewer bits by first encoding
the symbol counts.  To see why this is possible, observe that if the
counts for each symbol are known, then the items in the sequences are
not independent of each other.  If the size of the sequence at a given
node is $n$, and the number of items assigned to the right partition
is $n_r$, there are $\binom{n}{n_r}$ valid selections which is less
than the $2^n$ possible bitstrings of length $n$ describing the
comparisons.  Therefore, a description of the data that is more
efficient than the sample distribution entropy may be achieved by
encoding each node's decision bitvector using $\log_2(\binom{n}{n_r})$
bits.  In order to decode this, the sizes of the nodes must also be
recorded.  But the entire tree structure, along with the sizes for
each node, can be reconstructed from just the counts for the leaves
using the Fano coding algorithm~\cite{Cover}.

The symbol counts are not useful when data is uniform.  In that case,
the information is redundant, so including it makes random data more
expensive to encode.  Most data is random and incompressible (by any
criteria) \cite[p.41]{LiV08}, so the average performance suffers from
including the symbol counts.  However, the average extra cost is no
more than $M*log(N)$ bits, while the benefit can be dramatic for data
that is sparse or exhibits a lot of frequency regularity.

Encoding the partition decisions using the information theoretic limit
of $\log_2(\binom{n}{n_r})$ bits per node requires some way to map
between the sparse n-bit decision vectors and the dense
$\log_2(\binom{n}{n_r})$ bit descriptions.  For small values of $n$, a
lookup table can be constructed by some enumeration strategy.  As $n$
increases, this becomes infeasible, but a practical approximation is
to divide the sequence into windows of some size $W$, which is the
upper limit for the lookup table, and encode each of the
$\lceil{n/W}\rceil$ bitvectors using $\log_2(\binom{W}{n_{ri}})$ bits, with
$n_{ri}$ representing the number of ones in the $i$'th window.  This
requires also encoding the sequence $n_{ri}$, which can be done
efficiently because it sums to  $n_r$.

In addition to the record of decisions made at each node, the full
compression scheme needs to include a description of the symbols and
their counts.  The counts define a new, unordered, data sequence.
Because the sequence of counts is less complex than the starting
sequence, it can be encoded recursively using the same algorithm.  The
symbol values are ordered, so it is only necessary to encode their
differences.  Although these can be recursively compressed as well, if
the differences are large and non-repeating, they may not benefit from
additional compression iterations.  To ensure that the recursion
converges rapidly, a non-recursive, direct encoding should be used as
soon as the entropy of the data exceeds some threshold.

\bibliography{pivotcompress}
\bibliographystyle{alpha}

\end{document}